\documentclass[aps,twocolumn,superscriptaddress]{revtex4}
\usepackage{times}
\usepackage{color}
\usepackage{float}
\usepackage{epsfig}

\begin{document}

\title{{\it Landau damping} effects in the synchronization of conformist and contrarian oscillators}

\author{Tian Qiu}\thanks{Authors contributed equally to this work.}
\affiliation{Department of Physics, East China Normal University, Shanghai, 200241, China}

\author{Yue Zhang}\thanks{Authors contributed equally to this work.}
\affiliation{Department of Physics, East China Normal University, Shanghai, 200241, China}

\author{Jie Liu}\thanks{Authors contributed equally to this work.}
\affiliation{Department of Physics, East China Normal University, Shanghai, 200241, China}

\author{Hongjie Bi}\affiliation{Department of Physics, East China Normal University, Shanghai, 200241, China}

\author{S. Boccaletti}
\thanks{Corresponding author: stefano.boccaletti@gmail.com}
\affiliation{CNR-Institute of Complex Systems, Via Madonna del Piano, 10, 50019 Sesto Fiorentino, Florence, Italy}
\affiliation{The Embassy of Italy in Tel Aviv, 25 Hamered street, 68125 Tel Aviv, Israel}

\author{Zonghua Liu}
\affiliation{Department of Physics, East China Normal University, Shanghai, 200241, China}
\affiliation{State Key Laboratory of Theoretical Physics, Institute of Theoretical Physics, Chinese Academy of Sciences, Beijing 100190, China}

\author{Shuguang Guan}
\thanks{Corresponding author: guanshuguang@hotmail.com}
\affiliation{Department of Physics, East China Normal University, Shanghai, 200241, China}
\affiliation{State Key Laboratory of Theoretical Physics, Institute of Theoretical Physics, Chinese Academy of Sciences, Beijing 100190, China}

\date{\today}

\begin{abstract}
Two decades ago, a phenomenon resembling Landau damping was described in the synchronization of globally coupled oscillators: the evidence of a regime where the order parameter decays when linear theory predicts neutral stability for the incoherent state. We here show that such an effect is far more generic, as soon as phase oscillators couple to their mean field according to their natural frequencies, being then grouped into two distinct populations of {\it conformists} and {\it contrarians}. We report the analytical solution of this latter situation, which allows determining the critical coupling strength and the stability of
the incoherent state, together with extensive numerical simulations that fully support all theoretical predictions. The relevance of our results is discussed in relationship to collective phenomena occurring in polarized social systems.

\end{abstract}

\maketitle

%Introduction
In the forties, the Soviet physicist Lev Davidovich Landau (the 1962 Nobel Laureate for his theory on superfluidity) predicted the damping
(the exponential decrease as a function of time) of electrostatic
charge waves in a collision-less plasma \cite{Landau1946,Infeld1990}. After almost two decades of controversy,
the "Landau damping" (LD) was eventually verified experimentally \cite{Malmberg1964}, and it was even argued that similar phenomena could take place in galactic dynamics \cite{Bell1962}. Mathematically, LD is entirely due to
the occurrence of fake eigenvalues caused by analytic continuation, which lead to an exponential decay of the electric field even when the density perturbation does not.

More recently, Strogatz {\it et al.} investigated the synchronization transition in ensembles of coupled oscillators \cite{Strogatz1992},
and found a regime (below the synchronization threshold, and for which linear theory foretells neutral stability) where
the relaxation to the incoherent state is indeed exponential, with a decaying mechanism remarkably similar to LD.

In this Letter, we report on theoretical analysis and numerical simulations that demonstrate how the presence of LD is, actually, far more general in the synchronization route of globally coupled oscillators.
Without lack of generality, we start by considering a frequency-weighted Kuramoto \cite{Kuramoto1984} model of $N$ phase oscillators, in which units are coupled to the mean field according to their natural frequencies:
\begin{equation}\label{eq:model}
 \dot{\theta}_j=\omega_j + \frac{\kappa \omega_j }{N}\sum_{n=1}^{N}\sin(\theta_n-\theta_j),\quad j=1,...,N.
\end{equation}
Here $\theta_j$ ($\omega_j$) is the instantaneous phase (the natural frequency) of the $j$th oscillator, dot denotes a temporal derivative, and $\kappa>0$ is a global coupling strength parameter. The set $\{\omega_j\}$ of natural frequencies is drawn from a given frequency distribution (FD) $g(\omega)$ with, in general, both a positive and a negative domain.
Eq. (\ref{eq:model}) belongs to the class of the so-called generalized Kuramoto models \cite{Kuramoto1984,Strogatz1991,Strogatz2000,Bocca2002,Acebron2005,Arenas2008}. At variance with Refs. \cite{Zhang2013,Hu2014}, the effective coupling $\kappa\omega_j$ can be here either positive or negative (depending on the sign of the natural frequency $\omega_j$), reflecting the fact that the interactions among individuals can inherently be repulsive in real systems \cite{Zanette2005,Hong2011a,Hong2011b,Zhang2013a}. For instance, it is well known that both excitatory and inhibitory couplings characterize the interaction structure of neural ensembles \cite{Borgers2003,Qu2007}, as well as similar kinds of coupling can also be found in social interactions.

Eq. (\ref{eq:model}) is fully analytically solvable, and we here focus on unveiling the details of several novel phenomena of LD, which characterize the transition to synchronization in such systems. In its mean-field form, Eq. (\ref{eq:model}) can be written as
\begin{equation}\label{eq:mean-field}
\dot{\theta_j}=\omega_j+\kappa \omega_j r \sin(\phi-\theta_j),\quad j=1,...,N,
\end{equation}
where $r$ and $\phi$ are order parameters defined
by $r e^{i\phi} = \frac{1}{N} \sum_{j=1}^{N} e^{i\theta_j}$.
Here oscillators can be, in general, grouped into two populations, according to their effective coupling: those with positive $\omega_j$ will behave like {\it conformists} attempting to follow the global rhythm of the system, whereas those with negative $\omega_j$ will tend to act as {\it contrarians}, i.e. they will always try opposing the system`s global beat \cite{Hong2011a,Hong2011b,Zhang2013a}.
Let us then analyze the synchronization transition in Eq. (\ref{eq:model}) in the presence of a Lorentzian FD
$g(\omega) = (\gamma/\pi)/ [(\omega-\Delta) ^2 + \gamma ^2]$, where $\Delta$ and $\gamma$ are the central frequency and the width at half maximum, respectively, with $\Delta$ actually controlling the proportion of conformists to contrarians in the ensemble.

\begin{figure*}[htbp]
\begin{center}
\includegraphics[width=\textwidth]{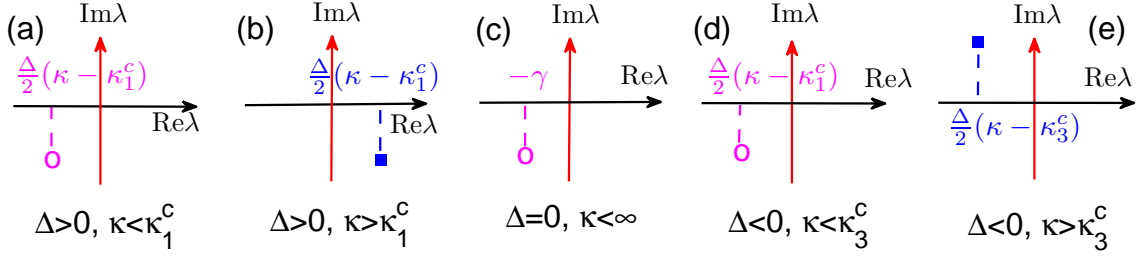}
\caption{(Color online) {\bf The spectra for the characteristic equation (\ref{eq:c})}.
For a  Lorentzian frequency distribution, the continuous spectrum is the whole imaginary axis. The solid squares denote, instead, the discrete eigenvalues. The purple ovals in (a), (c), and (d) mark the {\it ghost} (fake) eigenvalues predicted by the linear theory. As discussed in the text, such ghost eigenvalues remarkably control the decaying rate of order parameter $r(t)$ in the neutrally stable regimes, exactly as in the Landau damping context. It should be pointed out that the {\it ghost} eigenvalue in (d) could also be above the real axis, depending on parameters. For better visualization, we do not plot the other irrelevant (fake) eigenvalues.}\label{fig-eigen}
\end{center}
\end{figure*}

In the thermodynamic limit, i.e., $ N\to \infty $,  a density function $\rho(\theta,\omega,t)$ can be defined, where
$\rho \ d\theta$ denotes the fraction of oscillators with frequency $\omega$ whose phases have values between $\theta$ and  $\theta+d\theta$ at time $t$.
$\rho$ satisfies the normalization condition
$\int_0^{2\pi}\rho(\theta,\omega,t)d\theta=1$ for all $\omega$, and all $t$.
The evolution of $\rho$ is governed by the continuity equation
\begin{equation}\label{eq:continuity}
\frac{\partial\rho}{\partial t}+\frac{\partial (\rho\upsilon)}{\partial \theta}=0,
\end{equation}
where the velocity $\upsilon$ is given by
$\upsilon=\omega+\kappa \omega r \sin(\phi-\theta)$.
On its turn, the order parameter can be expressed as $re^{i\phi}
=\int_{0}^{2\pi}
\int_{-\infty}^{+\infty}e^{i\theta}\rho(\theta ,\omega ,t) g(\omega)d \omega d \theta$, and therefore
Eq. (\ref{eq:continuity}) can be rewritten as
\begin{eqnarray}\label{eq:continuity2}
\frac{\partial\rho}{\partial t}&=&-\frac{\partial}{\partial \theta}\{\rho[\omega+
\kappa\omega\int_0^{2\pi}
\int_{-\infty}^{+\infty}\sin(\theta-\theta^\prime)\cdot \nonumber \\
&&\rho(\theta^\prime,\omega^\prime,t)g(\omega^\prime)
d\omega^\prime d\theta^\prime ]\}.
\end{eqnarray}
For the incoherent state, $\rho_0(\theta,\omega,t)=1/(2\pi)$.
A small perturbation from that state, i.e.
\begin{equation}\label{eq:rho-expansion}
\rho(\theta,\omega,t)=\frac{1}{2\pi}+\epsilon [c(\omega,t)e^{i\theta}+
c^*(\omega,t)e^{-i\theta}+\eta^{\bot}(\theta,\omega,t)]
\end{equation}
can be considered, where $\epsilon\ll 1$, and $\eta^{\bot}(\theta,\omega,t)$ represents the higher Fourier harmonics.
Substituting Eq. (\ref{eq:rho-expansion})  into Eq. (\ref{eq:continuity2}), one gets the linearized characteristic equation
\begin{equation}\label{eq:c}
\frac{\partial c(\omega,t)}{\partial t}=-i\omega \cdot c(\omega,t)+\frac{\kappa \omega }{2}\int_{-\infty}^{+\infty}c(\omega^\prime,t)
g(\omega^\prime)d\omega^\prime.
\end{equation}

Eq. (\ref{eq:c}) has both continuous and discrete spectra (details form part of the material contained in the {\it SI}), and
one can extract the critical coupling strength $\kappa^c$ as well as determine the stability properties of the incoherent state
(which are schematically reported in Fig. \ref{fig-eigen}). Specifically:

(Case 1) $\Delta>0$: the conformists prevail over the contrarians.
$\kappa^c_1=2\gamma/\Delta$
and the spectra are schematically plotted in Fig. \ref{fig-eigen}(a) and (b). When $\kappa>\kappa^c_1$, Eq. (\ref{eq:c}) has a continuous spectrum  on the imaginary axis and a discrete eigenvalue in the right half complex plane. Accordingly, the incoherent state is unstable. When $0<\kappa<\kappa^c_1$, there is no discrete eigenvalue, and the incoherent state is neutrally stable.

(Case 2) $\Delta=0$: conformists and contrarians are equal in number. In this case, $\kappa^c_2=\infty$, implying that
synchronization can never be achieved. As shown in Fig. \ref{fig-eigen}(c), here no discrete eigenvalues exist outside the imaginary axis, and therefore the incoherent solution is always (i.e. for any arbitrary coupling strength) neutrally stable.

(Case 3) $\Delta<0$: the contrarians prevail over the conformists. $\kappa^c_3=-2\gamma/\Delta$, and the spectra are shown in Fig. \ref{fig-eigen}(d) and (e). When $\kappa>\kappa^c_3$ there is (besides the continuous spectrum on the imaginary axis) an eigenvalue in the left half complex plane. Thus, the incoherent state is linearly stable. When $0< \kappa <\kappa^c_3$, there is no discrete eigenvalue, and once again the incoherent state is neutrally stable.

The current situation shares connections and differences with the classical Kuramoto model \cite{Strogatz1991,Strogatz2000}.
For $\Delta>0$, the stability of the incoherent state is the same as that of Refs. \cite{Strogatz1991,Strogatz2000},
though the equations of the two models are essentially different. For $\Delta<0$, the incoherent state changes from neutrally stable to stable when $\kappa$ exceeds $\kappa^c_3$, and the bifurcation here only involves a change in stability of the state, while no new steady states emerge. Finally, for $\Delta=0$, the incoherent state is always neutrally stable, regardless of $\kappa$.
The two latter phenomena are novel, and inherent in our frequency-weighted Kuramoto model, which allows the two populations of conformist and contrarian oscillators to coexist.

In order to unveil LD in the model, we analytically extract the equation ruling
the relaxation behavior of $r(t)$  in all three neutrally stable regimes predicted by the linear theory, i.e., the cases corresponding to Figs. \ref{fig-eigen}(a), (c), and (d). As it will appear momentarily, the most remarkable finding is that $r(t)$ decays, indeed, exponentially in all cases.
The use of Eq. (\ref{eq:rho-expansion}) into  the definition of $r$ leads to  $r(t)=2\pi\epsilon|R(t)|$,
where
\begin{equation}\label{eq:Rt}
R(t)=\int_{-\infty}^{+\infty}c(t,\omega)g(\omega)d\omega.
\end{equation}
Then, Eq. (\ref{eq:c}) becomes
\begin{equation}\label{eq:c-R}
\frac{\partial c(\omega,t)}{\partial t}=-i\omega \cdot c(\omega,t)+\frac{\kappa \omega }{2}R(t).
\end{equation}
For any given initial condition $c_0(\omega)=c(0,\omega)$, the solution of Eq. (\ref{eq:c-R}) is:
\begin{equation}\label{eq:c-solution}
c(t,\omega)=e^{-i\omega t}[\int_0^t\frac{1}{2}\kappa\omega R(\tau) e^{i\omega \tau}d\tau +c_0(\omega)].
\end{equation}
Substitution of Eq. (\ref{eq:c-solution}) into Eq. (\ref{eq:Rt}) leads to
\begin{eqnarray}\label{eq:Rt2}
R(t)&=&\int_{-\infty}^{+\infty}e^{-i\omega t}
[\int_0^t\frac{1}{2}\kappa\omega R(\tau) e^{i\omega \tau} d\tau]g(\omega)d\omega \nonumber \\
&&+\int_{-\infty}^{+\infty} e^{-i\omega t} c_0(\omega)g(\omega)d\omega  \nonumber \\
&=&(\widehat{c_0g})(t)+\frac{1}{2}\kappa\int_0^t  \hat{G}(\tau)R(t-\tau)d\tau,
\end{eqnarray}
where the hat denotes the Fourier transform, i.e.,
$(\widehat{c_0g})(t)=\int_{-\infty}^{\infty} e^{-i\omega t} c_0(\omega)g(\omega)d\omega$ and $\hat{G}(t)=\int_{-\infty}^{\infty}e^{-i\omega t} \omega g(\omega)d\omega$.
The solution of Eq. (\ref{eq:Rt2}) is obtained by a Laplace transform (all details are reported in the {\it SI}),
and reads as
\begin{equation}\label{eq:mu}
R(t)=e^{\mu t}
=\exp\{[\frac{\Delta}{2}(\kappa-\frac{2\gamma}{\Delta})
-i( \frac{\gamma\kappa}{2}+\Delta)]t\}.
\end{equation}

\begin{figure}[h]
\begin{center}
\includegraphics[width=0.53\textwidth]{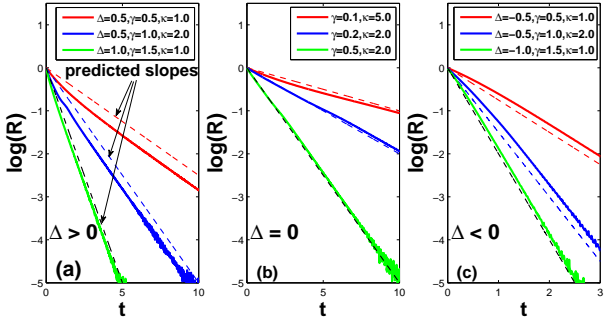}
\caption{(Color online) {\bf Landau damping effects in the order parameter $R(t)$}. Horizontal lines is time, solid lines refer to the numerical solutions of  Eq. (\ref{eq:model}), and dashed lines are the analytical predictions of Eq. (\ref{eq:mu}). The three log-linear plots refer to the cases $\Delta>0$ (a), $\Delta=0$ (b), and $\Delta<0$ (c). In the numerical simulations, the initial states of the system are set in the fully coherent states.    }\label{fig-mu}
\end{center}
\end{figure}

Notice that, in the derivation of Eq. (\ref{eq:mu}), we do not yet imposed specific constraints to the functional form of $\mu$,
therefore [based on the closed form of $R(t)$], we here below summarize what happens to the relaxation behavior of $r(t)$
in the three different cases of $\Delta$, while addressing at the same time the reader to the {\it SI} material for all technical details.
In the following, $\lambda_1$, $\lambda_2$, and $\lambda_3$ denote
the eigenvalues whose real parts are greater than, equal to, and less than 0, respectively.

(Case 1) $\Delta>0$. One obtains that the critical coupling strength is $\kappa^c_1=2\gamma/\Delta$, by setting Re$[\mu]$=0. When $\kappa<\kappa^c_1$ [where the linear theory predicts neutral stability, Fig. \ref{fig-eigen}(a)], $r(t)$ decays exponentially with Re$[\mu]$=$\Delta\kappa/2-\gamma$=Re$[\lambda_1]$. Therefore, it is remarkable that the damping rate is actually determined by the value of the {\it disappeared ghost} eigenvalue $\lambda_1$ in Fig. \ref{fig-eigen}(a), exactly like what happens in the LD mechanism.

\begin{figure}[htbp]
\begin{center}
\includegraphics[width=0.55\textwidth]{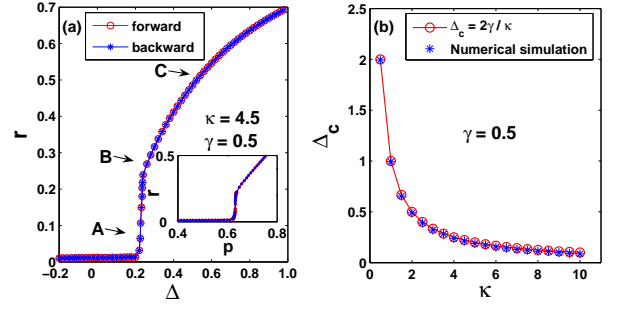}
\caption{(Color online) {\bf Synchronization transition in Eq. (\ref{eq:model}) in the presence of conformist and contrarian oscillators.}
(a) $r$ vs $\Delta$ (see text for definitions). Letters A, B, C denotes the three conditions that will be analyzed in the next Figure.
The inset reports $r$ vs $p$, i.e. the proportion of conformist oscillators in the ensemble, which (for any given $\gamma$) is entirely controlled by $\Delta$.
(b) Given $\gamma$, $\Delta_c$ comes out to be inversely proportional to $\kappa$. }\label{fig-r-Delta}
\end{center}
\end{figure}

(Case 2) $\Delta=0$. The incoherent state is  neutrally stable regardless of $\kappa$ [Fig. \ref{fig-eigen}(c)].
According to Eq. (\ref{eq:mu}), $r(t)$ {\it always} decays exponentially with Re$[\mu]=-\gamma$=Re$[\lambda_1]$, once again like if the {\it ghost} eigenvalue $\lambda_1$  in Fig. \ref{fig-eigen}(c) would still be present.

(Case 3) $\Delta<0$. From Eq. (\ref{eq:mu}), Re$[\mu]=\Delta\kappa/2-\gamma$, which is {\it always} negative. Therefore, $r(t)$ does decay exponentially for any value of $\kappa$, including in the neutrally stable regime predicted by linear theory.
Remarkably, the decaying rate Re$[\mu]=\Delta\kappa/2-\gamma$ is here (and must be) different from Re$[\lambda_3]=\Delta\kappa/2+\gamma=\frac{\Delta}{2}(\kappa-\kappa^c_3)$,  because the real part of $\lambda_3$ is positive when $\kappa<\kappa^c_3$.
Instead, the decaying rate Re$[\mu]$=Re$[\lambda_1]$, i.e.,
the {\it ghost} eigenvalue $\lambda_1$ actually controls
the decaying of $r(t)$, as shown
 in Fig. \ref{fig-eigen}(d)!

Finally, we compare our analytical predictions with direct numerical simulations of Eq. (\ref{eq:model}). To that purpose, we set $N=10,000$, and adopt a fourth-order Runge-Kutta method with integration step $\Delta t=0.01$. We first verify that the numerics match all theoretical predictions, as shown in Fig. \ref{fig-mu}, and confirm an exponential relaxation of $R(t)$ in the regimes of neutral stability, for all the three cases.
Next, starting from random initial conditions for the phases, we gradually step up (forward transition) and down (backward transition) $\Delta$ from (and to) a negative initial value
($\Delta=-0.2$), while keeping $\gamma=0.5$ and $\kappa=4.5$ constant, this way piecemeal modulating the proportion of conformist oscillators in the ensemble.
Results are shown in Fig. \ref{fig-r-Delta}(a), where it is seen that a continuous (fully reversible) synchronization transition occurs at the critical point $\Delta_c=2\gamma/\kappa$. In Fig. \ref{fig-r-Delta}(b),  $\Delta_c$ vs $\kappa$ is plotted, and numerical results perfectly verify the prediction
of an inverse proportion in the critical point. On its turn, this means that, as long as  the number of the conformists prevails over that of the contrarians, synchronization  will always occur for a large enough coupling strength.

\begin{figure}[h]
\begin{center}
\includegraphics[width=0.53\textwidth]{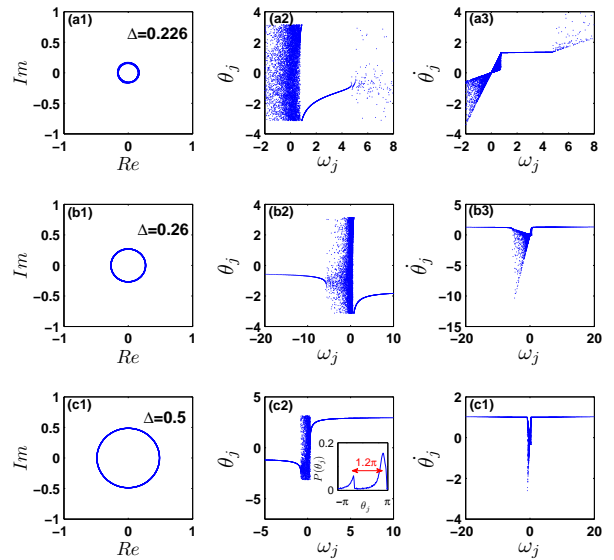}
\caption{(Color online) {\bf Characterization of the coherent states.} $\kappa=4.5 $ and $\gamma=0.5$.
Rows (a)-(c) correspond to $\Delta=0.226 $, $0.26$, and $0.5$, respectively, i.e. to the points A, B, and C of Fig. \ref{fig-r-Delta}(a).
Column 1 plots the order parameter in the complex plane after a transient stage, while columns 2-3 correspond to the snapshots of the distributions of the instantaneous phases and the frequencies at $t=2,500$.
In (a), the conformists slightly prevail over the contrarians, and only a small part of conformists form a coherent cluster, rotating at a certain frequency along the unit circle. In (b), the coherent cluster of conformists continuously expands, while the contrarians begin forming a coherent cluster. Both clusters rotate at the same frequency along the unit circle, with a constant phase difference between them.
In (c) more conformists are present in the system. The cluster of the conformists further enlarges, leading to the increase of the order parameter.
The inset in (c2) shows the phase distribution at that moment. The phase difference between two peaks is $1.2\pi$.}\label{fig-state1}
\end{center}
\end{figure}

Going further, one can even unveil the mechanisms and processes taking place during the path to synchronization. Namely, Fig. \ref{fig-state1} reports the instantaneous distributions of phases and frequencies that characterize the coherent states.  As $\Delta$ is below the critical value, synchronization cannot be achieved. When $\Delta$ exceeds $\Delta_c$, a phase-locking cluster of conformists ($\omega_i>0$) first appears [as shown in Fig. \ref{fig-state1}(a)], without an associated cluster of synchronized contrarians.  As  $\Delta$  increases, more and more conformists join the synchronized cluster. Only at an intermediate moment, the contrarians start forming a synchronized cluster, as shown in Fig. \ref{fig-state1}(b). This latter fact is actually striking, as it  demonstrates that in such ensembles the synchronization of contrarians can only be achieved {\it after} the synchronized cluster of conformists is large enough, as if the former process would be actually induced by the latter. By further increasing $\Delta$, the cluster of conformists becomes larger and larger by recruiting more and more drifting oscillators, which leads to larger order parameter, as shown in Fig. \ref{fig-state1}(c). The two clusters (that of conformist and that of contrarians) rotate with the same frequency along the unit circle. During the rotation, they are relatively static with each other, as shown in the inset of Fig. \ref{fig-state1}(c2), i.e. the two peaks in the phase distribution keep a constant difference $1.2\pi$ (or $0.8\pi$). As a consequence, the system enters (after synchronization) into a traveling wave state, which however (and generally) is not the $\pi$ state, i.e. that state where the phase difference between the two clusters is always $\pi$ \cite{Hong2011a}.

The occurrence of traveling wave states can be heuristically understood as follows. As pointed out in Ref. \cite{Zhang2014},
the phase-locking condition for a pair of oscillators in system (\ref{eq:model}) is
$\Delta\omega_{ij}=|\omega_j-\omega_i|/(|\omega_j|+|\omega_i|)<\kappa r$, i.e. a pair of oscillators is forbidden to synchronize with each other (for any given $\kappa$) if such a condition does not hold. Now, if a pair of oscillators have natural frequencies with different signs, $\Delta\omega_{ij}=1$. In the incoherent state,  $\kappa r$ is typically much less than 1, and therefore those oscillator pairs with $\Delta\omega_{ij}=1$ will most likely violate the above condition. At variance, oscillator pairs with the same frequency sign are easier to synchronize, especially when their natural frequencies are close enough. This explains why clusters always form among conformists or contrarians, as observed in Fig. \ref{fig-state1}.
In the traveling wave state, both coherent clusters of conformists and contrarians rotate with the same instantaneous frequency that is greater than 0.
According to Eq. (\ref{eq:mean-field}), the conformists turn to approach the mean-field phase $\phi$ due to $\omega_i>0$.
Nevertheless, in order to make the instantaneous frequency $\dot{\theta}_i>0$ of contrarians, $\sin(\phi-\theta)$ must be less than 0, i.e. their phases will always rotate against the mean-field phase $\phi$, and one will observe a phase difference between the two clusters (conformists and contrarians) which is always greater than $\pi/2$.

In summary, we investigated the synchronized dynamics of an ensemble of phase oscillators when the coupling to the mean field is frequency-weighted and two populations of oscillators (conformists and contrarians) can be identified. Analytically, we derived the critical coupling strength for synchronization and determined the stability of the incoherent state. In all regimes where the linear theory predicts neutral stability, the order parameter decays exponentially, in analogy with the Landau damping effect in plasma physics. Extensive numerical simulations fully support the theoretical predictions, and show that a continuous synchronization transition occurs by changing the ratio of conformists and contrarians. Traveling wave states are generically observed after synchronization.
Together with providing generic evidence of Landau damping effects in the synchronization of coupled oscillators, our results are of significance in that they contribute to shed light on the mechanisms at the basis of some phenomena beheld in social sciences. In many polarized social systems (such as two parties in the congress of a country, or fans of two derby football clubs) contrarians always confront with conformists. In these situations, a phenomenon is frequently observed, i.e. the weaker side becomes more united as the stronger sides becomes more powerful, while the two sides never compromise with each other. Our results may then enlighten the reasons for the occurrence of such circumstances.

Work partially supported by
the National Natural
Science Foundation of China under Grants
No.  11375066  and No. 11135001; and the Open Project Program of State Key Laboratory of Theoretical Physics, Institute of Theoretical Physics, Chinese Academy of Sciences, China (No. Y4KF151CJ1).


\begin{thebibliography}{99}

%Landau damping
\bibitem{Landau1946}
L. Landau, J. Phys. USSR {\bf 10}, 25 (1946).

\bibitem{Infeld1990}
E. Infeld and G. Rowlands, {\it Nonlinear Waves, Solitons and Chaos}
(Cambridge Univ. Press, New York, 1990).

\bibitem{Malmberg1964}
J. H. Malmberg and C. B. Wharton,
Phys. Rev. Lett. {\bf 13}, 184 (1964).

\bibitem{Bell1962}
D. Lynden-Bell,  Mon. Not. R. Astr. Soc. {\bf 124}, 279 (1962).

\bibitem{Strogatz1992}
S. H. Strogatz,  R. E. Mirollo,  and  P. C. Matthews,
Phys. Rev. Lett. {\bf 68}, 2730 (1992).

\bibitem{Kuramoto1984}
Kuramoto, Y. {\it Chemical Oscillations, Waves, and Turbulence.} (Springer, New York, 1984).


\bibitem{Strogatz1991}
S. H. Strogatz and R. E. J.  Mirrolo,  J. Stat. Phys. {\bf 63}, 613 (1991).

\bibitem{Strogatz2000}
S. H. Strogatz, Physica D {\bf 143}, 1 (2000).

\bibitem{Bocca2002}
S. Boccaletti, J. Kurths, G. Osipov, D.L. Valladares and C.S. Zhou, Phys. Rep. {\bf 366}, 1 (2002).

\bibitem{Acebron2005}
 J. A. Acebr{\'o}n,  L. L. Bonilla,   C. J. P{\'e}rez Vicente,   F. Ritort,  and R. Spigler, Rev. Mod. Phys. {\bf 77}, 137 (2005).


\bibitem{Arenas2008}
A. Arenas, A.  D{\'i}az-Gilera,  J. Kurths, Y.  Moreno,  and C.  Zhou, Phys. Rep. {\bf 469}, 93 (2008).

\bibitem{Zhang2013}
X. Zhang,  X.  Hu,  J.  Kurths,  and  Z. Liu,  Phys. Rev. E {\bf 88}, 010802(R) (2013).

\bibitem{Hu2014}
Hu, X.  {\it et al.}, Sci. Rep. {\bf 4}, 7262 (2014).


%Conformist-contrarian
\bibitem{Zanette2005}
 D. H. Zanette,  Europhys. Lett. {\bf 72}, 190 (2005).

\bibitem{Hong2011a}
H. Hong and S. H.  Strogatz, Phys. Rev. Lett. {\bf 106}, 054102 (2011).

\bibitem{Hong2011b}
H. Hong and S. H.  Strogatz,  Phys. Rev. E. {\bf 84}, 046202 (2011).

\bibitem{Zhang2013a}
X. Zhang, Z. Ruan, and Z. Liu, Chaos {\bf 23}, 033135 (2013).

%biological
\bibitem{Borgers2003}
C. B{\"o}rgers and N. Kopell, Neural Comput. {\bf 15}, 509 (2003).

\bibitem{Qu2007}
 Z. Qu, Y.  Shiferaw,  and J. N. Weiss, Phys. Rev. E. {\bf 75}, 011927 (2007).

\bibitem{Zhang2014}
X. Zhang,  Y.  Zou, S. Boccaletti,  and  Z. Liu, Sci. Rep. {\bf 4}, 5200 (2014).

\end{thebibliography}
\end{document}